\documentclass[printer]{aa}  

\usepackage{graphicx}
\usepackage{txfonts}
\usepackage{natbib,twoopt}
\usepackage[breaklinks=false]{hyperref} 
\bibpunct{(}{)}{;}{a}{}{,}             
\usepackage{color}

\begin{document}

\title{Sizes and albedos of Mars-crossing\\ asteroids from WISE/NEOWISE data}


\author{V. Al\'i-Lagoa
  \inst{1,2}
  \and
  M. Delbo'\inst{1}
}

\institute{
  Laboratoire Lagrange, Universit\'e C\^ote d'Azur, Observatoire de la C\^ote d'Azur, CNRS\\
  Blvd de l'Observatoire, CS 34229, 06304 Nice cedex 4, France\\
  \and
  Max-Planck-Institut f{\"u}r extraterrestrische Physik, Giessenbachstrasse 1, 85748 Garching, Germany\\
  \email{vali@mpe.mpg.de}
}

\date{Received ; accepted February 28, 2017}


\abstract{
  Mars-crossing asteroids (MCs) are a dynamically unstable group between the main belt and the near-Earth populations. Characterising the physical properties of a large sample of MCs can help to understand the original sources of many near-Earth asteroids, some of which may produce meteorites on Earth. 
}
{
  Our aim is to provide diameters and albedos of MCs with available WISE/NEOWISE data. 
}
{
  We used the near-Earth asteroid thermal model to find the best-fitting values of equivalent diameter and, whenever possible, the infrared beaming parameter. With the diameter and tabulated asteroid absolute magnitudes we also computed the visible geometric albedos. 
}
{
  We determined the diameters and beaming parameters of 404 objects observed during the fully cryogenic phase of the WISE mission, most of which have not been published elsewhere. We also obtained 1572 diameters from data from the 3-Band and posterior non-cryogenic phases using a default value of beaming parameter.  
  The average beaming parameter is 1.2$\pm$0.2 for objects smaller than 10 km, which constitute most of our sample. This is higher than the typical value of 1.0 found for the whole main belt and is possibly related to the fact that WISE is able to observe many more small objects at shorter heliocentric distances, i.e. at higher phase angles. We argue that this is a better default value for modelling Mars-crossing asteroids from the WISE/NEOWISE catalogue and discuss the effects of this choice on the diameter and albedo distributions. 
  
  We find a double-peaked distribution for the visible geometric albedos, which is expected since this population is compositionally diverse and includes objects in the major spectral complexes. However, the distribution of beaming parameters is homogeneous for both low- and high-albedo objects. 
}
{}


\titlerunning{Sizes and albedos of Mars-crossing asteroids}

\maketitle

\section{Introduction}

Mars-crossing asteroids (MCs) occupy unstable orbits between the main belt and
the population of near-Earth asteroids (NEAs). 
Knowledge about their diameters and albedos can be combined with dynamical
studies
\citep[see e.g. ][]{Migliorini1998,Michel2000,Morbidelli2002,Granvik2016} to
link many NEAs with their source regions in the asteroid belt. 
This knowledge, combined with the physical properties of NEAs as a population,
can lead to better estimates of crucial parameters required to assess impact
risk and select adequate mitigation strategies in case a collision with an NEA
is deemed probable. Risk assessment and mitigation are major areas of interest
of the NEOshield-2 project \citep{Harris2013}, which focuses on small-sized
NEAs ($<$1~km), the least accessible for observation. As part of this effort,
in this article we provide diameters and visible geometric albedos of
Mars-crossing asteroids observed by the Wide-field Infrared Survey Explorer
(WISE) in the different phases of the survey \citep{Wright2010}. 
Ultimately, these can also help the community expand the scientific return
gained from spacecraft missions to NEAs by providing context for everything we
learn about the individual asteroids visited
\citep[e.g. ][]{Michel2010,Bottke2015,Lauretta2015}. 

We took the definition of MC given in the Jet Propulsion Lab Small-Body Database Search Engine\footnote{See http://ssd.jpl.nasa.gov/sbdb\_query.cgi}, i.e. all objects with semi-major axis $a < 3.2$ au and perihelion distance ($q$) in the range $1.3~\mathrm{au}< q <1.666~\mathrm{au}$. Other authors may apply different criteria to classify MCs \citep[e.g.][]{Michel2000} but we opted for the broadest definition possible. 

To compute diameters we modelled infrared (IR) data obtained by the WISE/NEOWISE mission \citep{Mainzer2011a,Mainzer2014}. More details and references about the data catalogue and WISE are given in Sect.~\ref{sec:data}. 
We queried the IRSA/IPAC catalogues for any available data of all objects in the MC list. We used the near-Earth asteroid thermal model (NEATM) of \citet{Harris1998} to find what values of diameter ($D$) best fit those data (see Sect.~\ref{sec:model}). Although the NEATM was conceived to produce better results for NEAs, 
 it has been successfully applied to all small bodies without atmospheres, including cometary nuclei and trans-Neptunian objects \citep[e.g. ][]{Fernandez2013,Santos-Sanz2012,Bauer2013}.

We determined the diameters and beaming parameters of 404 objects observed during the fully cryogenic phase of the WISE mission, most of which have not been published elsewhere. With a suitable default value of $\eta$, we also obtained 1572 more diameters from data from the 3-Band and posterior non-cryogenic phases. Combining the asteroid absolute magnitudes tabulated by the Minor Planet Center with these diameters we also computed the visible geometric albedos. In the absence of any other information, visible albedos can help distinguish between typically low-albedo primitive asteroids, spectrally associated with carbonaceous chondrites, and higher-albedo types associated with ordinary chondrites and other more processed meteorites that contain a smaller amount of volatiles than carbonaceous chondrites \citep[for a review and pertinent caveats, see][]{DeMeo2015}.

To compare our diameters with previously published values, we searched for objects in the MCs list in the catalogues of \citet{Masiero2011,Masiero2013,Masiero2014} and \citet{Nugent2015,Nugent2016}. We found a total of 48 and 534 objects, respectively, with which we compared our sizes and albedos on a one-to-one basis. In Sect.~\ref{sec:comparison} we show that for equal input parameters we obtain small systematic deviations between our sizes and albedos but that these fall well within the minimum expected errors of the NEATM \citep{Harris2006,Mainzer2011b}. 

The most salient features of the beaming parameter and albedo distributions of our sample are presented in Secs.~\ref{sec:etas} and \ref{sec:albedos}. In Sect.~\ref{sec:uncertainties} we discuss to what extent the particular choice of the default beaming parameter can bias NEATM diameters and the corresponding values of visible geometric albedo and justify our choice of default beaming parameter for the MCs.

\section{WISE/NEOWISE data\label{sec:data}}

The WISE survey provided measurements in up to four bands in the short-wavelength and thermal IR, W1$=$3.4~$\mu$m, W2$=$4.6~$\mu$m, W3$=$12~$\mu$m, and W4$=$22~$\mu$m \citep[see][and references therein]{Wright2010} during the fully cryogenic stage of the mission (see below). Enhancements to the WISE data processing system were designed by what is collectively referred to as the NEOWISE project to allow detection and archiving of solar system objects \citep{Mainzer2011a}. 
\citet{Nugent2016} provide an updated compendium of the different works of the NEOWISE team reporting asteroid sizes, albedos and infrared beaming parameters derived from NEOWISE data.

The survey scan angular velocity of WISE was such that an inertial source was observed about twelve times per apparition, once every $\sim$1.58 hours \citep{Cutri2012}. Owing to their non-sidereal motion, asteroids were sometimes not exposed on the following scan but at a subsequent pass, usually 3.16 hours later. The fully cryogenic phase of the mission covered 120\% of the sky\footnote{http://irsa.ipac.caltech.edu/Missions/wise.html.}. Asteroid observations were obtained in the four bands but mostly in bands W3 and W4, where these bodies emit a great fraction of their thermal radiation (see Sect.~\ref{sec:model}). After one of the coolant tanks ran out, 30\% of the sky was surveyed in bands W1, W2, and W3, (3-Band phase) and once the cooling system became inoperative (Post-Cryo survey), an additional 70\% of the sky was surveyed in bands W1 and W2. The mission is currently operational and collecting W1 and W2 data since its reactivation in December 2013 after a hiatus of 32 months \citep{Mainzer2011a,Mainzer2014}. Two NEOWISE Reactivation releases, covering two years of observations, have been published so far.

The procedure we followed to download WISE/NEOWISE asteroid data and reject low-quality and/or contaminated exposures is based on a combination of criteria taken from several works by the NEOWISE team \citep{Wright2010,Masiero2011,Mainzer2011a,Mainzer2011b,Grav2012,Cutri2012}. We downloaded the reported observation tracklets from the Minor Planet Center (MPC) and used a cone-search radius of 1'' around these coordinates when we queried the IRSA/IPAC catalogues; we required the tracklet date to be within 4 seconds of the catalogue date. We rejected data with magnitudes brighter than the point-source saturation thresholds given by \citet{Cutri2012}, which means that we do not use partially saturated data. We also rejected any catalogue entry with a magnitude error bar greater than 0.25 magnitudes and artefact flag other than 0, p, or P\footnote{A value of ``0'' indicates no known artefact affecting the exposure, whereas ``P'' and ``p'' indicate possible spurious detections of and contamination by a latent image, respectively. These flags were found to be overly conservative and they have been widely used in previous works \citep[e.g. ][]{Masiero2011}.}. Finally, we queried the All-sky catalogue with a cone-search radius of 6.5'' around the tracklets to identify potentially contaminating inertial sources. When matches were found, we accepted the asteroid fluxes only if the inertial sources were much fainter, namely if their reported flux was $\leq$5\% of the asteroid fluxes.

\section{Near-Earth asteroid thermal model (NEATM)\label{sec:model}}

The basic assumption of some asteroid thermal models is that the tempearure ($T$) of a surface element reaches thermal equilibrium  instantaneously with the incident solar energy. Since no a priori information about the asteroids is usually available, a non-rotating spherical shape with a smooth surface and no heat conduction towards the subsurface is assumed. This means that the instantaneous temperature of a surface element will be determined only by how much energy it absorbs. If we assume black-body emission, we then have 
\begin{equation}
  \mu(1-A)\frac{S_\odot}{r^2}=\eta\epsilon\sigma T^4,
  \label{eq:energy}
\end{equation}
where $\mu$ is the cosine of the angle between the element's normal and the direction towards the sun, $A$ the bolometric Bond albedo, $S_\odot$ the solar incident energy at 1 au, $r$ the heliocentric distance, $\epsilon$ the emissivity \citep[assumed to be 0.9; see][and references therein]{Delbo2015}, and $\sigma$ the Stefan-Boltzmann constant. The factor $\eta$, the infrared beaming parameter, is introduced to account for the non-linear increase in the thermal flux observed at low phase angles, i.e. the angle subtended by the observer and the sun from the asteroid's point of view. The standard thermal model used a value of $\eta=0.756$, which resulted in a better match to the sizes of large asteroids determined from occulations \citep[see][for a review]{Lebofsky1989}. \citet{Harris1998} used $\eta$ as a free parameter to improve the radiometric sizes for NEAs, for which the standard thermal model produced systematically smaller values \citep{Veeder1989}. The treatment of $\eta$ as a free parameter has the benefit of partially compensating for the simplifying assumptions of the model---after all, asteroids are not spherical, they rotate, and their surfaces are non-Lambertian.

Equation~\ref{eq:energy} allows us to compute the temperatures of all surface elements of the sphere at the time of the observations. 
Our shape model is a sphere with a diameter of 1 km or, more precisely, a polyhedron composed of 2296 triangular facets of equal area. Thus, the model flux corresponding to a body with equivalent diameter $D$~km would be the sum of all fluxes coming from all illuminated facets of said sphere visible to the observer multiplied by $D^2$. Non-illuminated facets are neglected. To account for the geometry of the observations, the model fluxes are also scaled by the inverse of the square of the distance between the asteroid and WISE (for simplicity, we take the geocentric distance since WISE is in a low Earth orbit). 

We follow the NEATM implementation of the NEOWISE team closely \citep{Masiero2011,Masiero2014,Mainzer2011b}, but with two main differences. First, whenever possible, we only model data dominated by thermal emission (wavelengths longer than $\sim$7 $\mu$m), i.e. W3 and W4 data. This avoids influencing the NEATM with the introduction of additional model assumptions needed to account for the reflected light component that usually becomes relevant at wavelengths shorter than $~$7 $\mu$m, which affects W1 and W2 data. Secondly, the visible geometric albedo ($p_V$) is not a parameter of our model. It is computed from the fitted diameter and the asteroid absolute magnitudes ($H$) tabulated by the MPC (as of August 2016). We used the expression
\begin{equation}
  p_V = \left(\frac{1329~\mathrm{km}}{D}10^{-H/5}\right)^2.
  \label{eq:albedo}
  \end{equation}  
Absolute magnitudes are constantly updated in the MPC, new $p_V$-values can be easily recomputed from our radiometric diameters and the updated $H$-values from Eq.~\ref{eq:albedo}. Alternatively, $H$ magnitudes from other catalogues could also be used \citep{Muinonen2010,Oszkiewicz2011,Veres2015,Williams2012}.

The $H$-values are also used along with the slope parameters ($G$) of \citet{Bowell1989} as input for the NEATM to obtain an estimate of the Bond albedo ($A$), which is required but not known a priori. We start with an initial guess value of $p_V=0.10$ and compute the corresponding diameter (from Eq.~\ref{eq:albedo}) and Bond albedo \citep[e.g. from $G$ and Eqs.~5 and 6 in][]{Lebofsky1989}. We run the model and find the best-fitting value of $D$, which we use to recompute $A$. We iterate until we arrive at a desired level of convergence for the Bond albedo, normally requiring four or five iterations to achieve 1\%. This procedure ensures that the Bond albedo is consistent with the size and the $H$ and $G$ values \citep[see, e.g. ][]{Delbo2004,Mueller2007}. On the other hand, we note that our diameters are somewhat insenstive to $H$ and $G$ as long as at least one band dominated by purely thermal data is available, because temperatures do not change significantly over the relatively narrow range of asteroidal Bond albedos (from Eq.~\ref{eq:energy}, $T \propto (1-A)^{\frac{1}{4}}$).

Whether we fit $\eta$ along with the diameter or not depends on the availability of data in the WISE bands in each case. We consider three possibilities:
\begin{enumerate}
\item The diameter and $\eta$ are both fitted only in cases in which there are data available in both of the purely thermal bands, W3 and W4. We model separately groups of data taken more than three days apart \citep[e.g.][]{Mainzer2011b} because $\eta$ is not a purely physical property of the asteroid surfaces and there is no unique $\eta$-value for a given asteroid. For example, it can vary depending on the phase angle of the observations ($\alpha$) or the particular aspect angle of the direction towards the observer, which can change the visible projected area of an irregular object significantly \citep[see e.g. ][]{Delbo2007}. 
\item When data are available in only one purely thermal bandpass (either W3 or W4, but not both), we model them assuming a default value of $\eta_d=1.2$ based on the average value obtained in this work for this sample of MCs (see Sect.~\ref{sec:etas} below). This default value is different from that used by NEOWISE team, and we discuss the effects of this choice on the diameters and albedos in Sect.~\ref{sec:uncertainties}. 
\item  If W3 and W4 data are not available but W2 data are, we need to account for the contribution of the reflected light, since fluxes at 4.6~$\mu$m are usually a mixture of emitted and reflected sunlight in different proportions depending on the heliocentric distance and albedo. We assume a default value for the albedo at 4.6~$\mu$m of $p_{\mathrm{W2}}=1.4\times p_V$, based on a compromise between the more neutral spectra of C-complex asteroids and the higher spectral slopes characteristic of many of the S-complex spectral classes \citep[e.g.][]{Masiero2014}. We also adopt the default value of $\eta_d=1.2$.
\end{enumerate}
Finally, following \citet{Nugent2016}, in the few cases when the resulting $p_V$ value was too low to be physical---we set the limit to 0.025---we re-ran the model with a lower fixed value of $\eta_d=1.0$, and reiterated with $\eta_d=0.9$ if we still did not obtain values of $p_V$ higher than the minimum values.

\subsection{Comparison with the NEOWISE team diameters and albedos\label{sec:comparison}}

To compare our results with those of the NEOWISE team and validate our model, we took the list of MCs provided by the JPL Horizons search engine and found 47 MCs in Table 1 of \citet{Masiero2011}, observed during the full cryogenic phase of the WISE mission, and 532 in \citet{Nugent2015} and \citet{Nugent2016}, observed during the reactivation phase. 

One source of discrepancy in the albedos is the different sources from which we obtained the $H$ values. \citet{Pravec2012} obtained $H$ magnitudes for nearly 600 asteroids and compared them to those tabulated in the most widely quoted catalogues, including the MPC. They found the latter to be systematically overestimated (the MPC $H$-values are greater than theirs) for objects smaller than $H>12$ and more importantly $H>14$, which presented an average offset of -0.5 mag). For this reason, \citet{Nugent2016} used an improved catalogue of $H$-values by \citet{Williams2012}, in which said offset is corrected. 
To give an idea of how much our results could be affected by this, we computed the differences in $H$-values for the objects featured in our catalogue and/or in \citet{Nugent2015} and \citet{Nugent2016}. Out of 489 objects, 200 have the same $H$ value, 186 are brighter in our catalogue by -0.25 mag on average, and 103 are about 0.1 mag fainter on average. Although these systematic deviations are comparable to the expected error bars of the absolute magnitudes (typically 0.1 mag, but not infrequently as large as 0.3 mag), some albedo values can be biased. Thus, to compare our results (see below), we only consider those cases with equal $H$ magnitudes. 

We computed relative differences between our diameters and the Masiero et al. values only in those cases where the absolute magnitudes were the same and the beaming parameter was either fitted (6 values) or not fitted (21 values) in both catalogues. Our sizes are 2.3\% smaller on average and our albedos 3\% higher.

\begin{figure}
  \begin{center}
    \includegraphics[width=0.7\hsize]{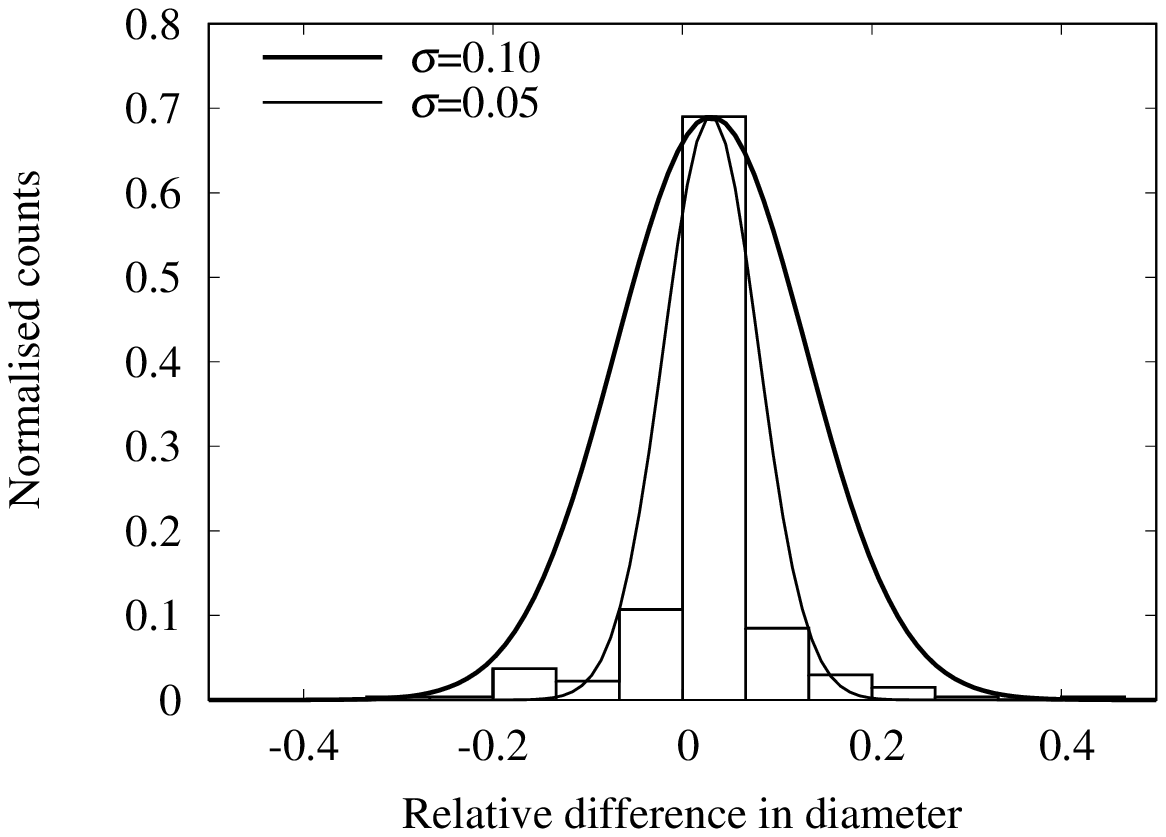}
    
    \includegraphics[width=0.7\hsize]{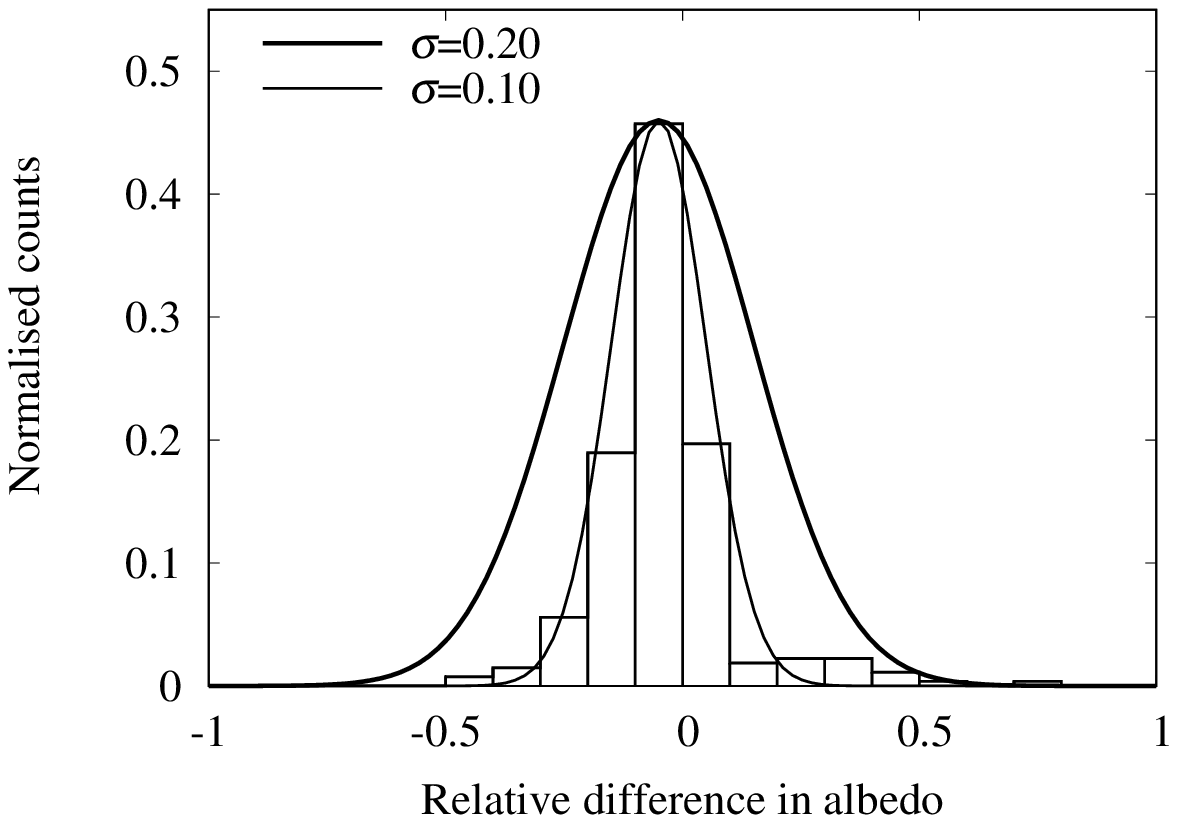}
    \caption{
      Relative differences between our diameters (upper panel) and visible geometric albedos (lower panel) and those reported in \citet{Nugent2015,Nugent2016}. We used the same default value of beaming parameter and 4.6-$\mu$m albedos that they used. The curves are Gaussian functions with the same mean values but different standard deviations. Our diameters are 3\% higher on average and our albedos 3\% lower.
      \label{fig:hist_comparison}
    }
  \end{center}
\end{figure}

We also modelled the Nugent et al. objects using their default values of beaming parameter and (W2) 4.6-$\mu$m albedo, 0.95 and 1.5$\times p_V$, respectively. Figure~\ref{fig:hist_comparison} shows the histogram of the relative difference between their diameters and ours for all objects which had the same $H$ values in both catalogues (271 entries). For comparison, we also plotted Gaussian functions with the same average value and standard deviations of 0.05 and 0.10. Similarly, the lower panel shows the same for the visible geometric albedo values, for which we expect a broader distribution since $p_V\propto D^{-2}$ (Eq.~\ref{eq:albedo}). In this case, our sizes are 3\% higher on average and our albedos 3\% lower.

These systematic differences fall well within the typical diameter error bars  quoted for the NEATM. Figure~4 in \citet{Harris2006} shows, for example, that the fractional error in NEATM diameters fitted to synthetic fluxes generated by a spherical shape with non-zero thermal inertia can range from $<$1\% for low thermal inertia and low phase angles, to 15\% for high thermal inertia and high phase angle. The error for non-spherical shapes has not been quantified in this manner, but the expectation is that the discrepancies would be larger. Therefore, it seems reasonable to assume a minimum error bar of 10\% for the diameter \citep[in agreement with the estimate of][]{Mainzer2011b}; we note, however, that the model can be much less accurate in some particular cases, for example when very few data points are available for the fit (see Secs.~\ref{sec:etas} and \ref{sec:uncertainties}). 
 
\section{Results\label{sec:results}}

Table~\ref{table:all} provides our best-fitting values of parameters and other relevant information regarding the observations and input parameters. Unlike in our previous works, we do not impose an absolute minimum number of data points to produce a fit. However, for a given group of observations not separated by more than three days, we do reject all data in a band if it does not contain at least 40\% of the data with the highest number of valid detections \citep{Mainzer2011b}. This means that in Table 1 there are beaming parameter values derived from one measurement in W3 and one in W4, but never from one measurement in W3 and three in W4, for example; in other words, the number of reliable detections needs to be reasonably similar in both bands.

In the following sections we show and discuss some noteworthy features of the beaming parameter and albedo distributions of our sample. 
\begin{table*}[!]
  \begin{center}
    \caption{Best-fitting values of size ($D$) and beaming parameter ($\eta$) and corresponding visible geometric albedos ($p_V$) for all our objects.
      \label{table:all} 
    }             
    \scriptsize    
    \begin{tabular}{r r r r c c c c c  c c c c}
      \hline\hline                 
      Object &  $H$ & $G$ & $D$ (km) & $p_V$  & $\eta$ & n$_\mathrm{W2}$ & n$_\mathrm{W3}$ & n$_\mathrm{W4}$ & $r$ (au) & $\Delta$ (au) & $\alpha$ (degree) & MJD \\
      \hline         
      00132  &   9.38  &  0.15  &  50.13 & 0.124 &  0.92 &    0 & 15 & 15 & 3.30812 & 3.14966 & 17.88688 & 55395.8002 \\ 
      00323  &   9.73  &  0.15  &  29.23 & 0.265 &  0.94 &    0 & 15 & 15 & 2.77473 & 2.50662 & 21.25193 & 55321.6229 \\ 
      00391  &  10.80  &  0.15  &  17.33 & 0.282 &  0.91 &    0 & 11 & 11 & 2.76976 & 2.58080 & 20.89958 & 55244.0457 \\ 
      00391  &  10.80  &  0.15  &  19.66 & 0.219 &  1.01 &    0 & 12 & 12 & 2.28260 & 1.96104 & 26.29792 & 55409.9965 \\ 
      00512  &  10.68  &  0.15  &  18.70 & 0.270 &  1.03 &    0 & 14 & 14 & 2.18148 & 1.94928 & 27.23824 & 55284.5366 \\ 
      00699  &  11.72  &  0.15  &  12.19 & 0.244 &  1.02 &    0 & 14 & 13 & 3.66612 & 3.43622 & 15.92318 & 55325.3199 \\
      01009  &  13.90  &  0.15  &   6.47 & 0.116 &  1.26 &    0 &  7 &  5 & 3.04221 & 2.85917 & 18.86819 & 55217.0517 \\ 
      01011  &  12.74  &  0.15  &   7.56 & 0.248 &  1.19 &    0 &  9 &  9 & 2.56987 & 2.36658 & 22.80497 & 55276.0682 \\ 
      01131  &  12.90  &  0.15  &   6.53 & 0.287 &  1.13 &    0 & 16 & 16 & 2.17259 & 1.83759 & 26.90384 & 55243.0204 \\ 
      01139  &  12.51  &  0.15  &   8.24 & 0.258 &  0.90 &    0 & 22 & 11 & 2.24136 & 1.97573 & 26.69370 & 55285.0652 \\ 
      \hline     
    \end{tabular}
    \tablefoot{
      Absolute magnitude and slope parameters ($H$, $G$) were retrieved from the Minor Planet Center.
      W1 through W4 indicate the number of observations used in each WISE band. 
      When $\eta$ could not be fitted, we show the negative value of the default $\eta$ that we used. 
      We took the geometry of observation for each epoch from the Miriade Ephemeris Generator. ``MJD'' refers to the modified Julian Date of the first observation, whereas $r$, $\Delta$ and $\alpha$ are the average values of heliocentric and geocentric distances and phase angle of each group of epochs.  
      Minimum relative errors of 10\%, 15\%, and 20\% should be considered for $D$, $\eta$, and $p_V$ when $\eta$ could be fitted  to at least five measurements in bands W3 and W4. Otherwise, the minimum errors for the diameters would be 20\% and 40\% for the albedos \citep[see][]{Nugent2016}.  
      Here we provide a small set of results to illustrate the format of the full table, which can be downloaded from the electronic version of the article.  
    }
  \end{center}
\end{table*}

\subsection{Infrared beaming parameters\label{sec:etas}}

\begin{figure}
  \begin{center}
    \includegraphics[width=0.70\hsize]{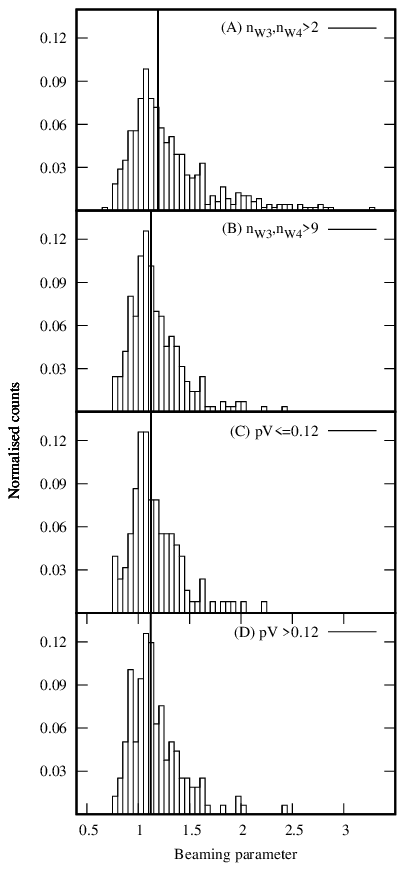}    
    \caption{
      Histograms of infrared beaming parameter of MCs observed during the full cryogenic phase of the WISE mission. (A) All objects with at least three data points in each one of the thermal bands. (B) All objects with at least ten data points. (C) Low- and (D) high-albedo objects from the sample of panel B.
      \label{fig:hist_eta}
    }
  \end{center}
\end{figure}

Figure~\ref{fig:hist_eta} shows normalised histograms for four different samples. The first two, labelled A and B, include all objects with $\eta$ fitted to at least three (n$_\mathrm{W3}$,n$_\mathrm{W4}>$2) and ten (n$_\mathrm{W3}$,n$_\mathrm{W4}>$9) measurements in each of the purely thermal bands, W3 and W4. Imposing these two different criteria to fit $\eta$ illustrates how increasing the minimum number of detections per band can be used to judge the fit reliability (this is discussed further below) and can help identify questionable extreme values \citep[see Appendix~A in][]{Ali-Lagoa2016}. 

The Kolmogorov-Smirnof (KS) test rules out that samples A and B are compatible with a $p$-value of $\sim 0.001$. We obtain $p$-values greater than 1 per cent only for samples with at least five data points in each thermal band, so we take this as a criterion to ensure reliable fits with uncertainties in diameter close to, but not lower than, 10\%. With this criterion, the sample's average beaming parameter is 1.2$\pm$ 0.2 (median 1.1$\pm$ 0.2). Based on the analysis of Sect.~\ref{sec:uncertainties} we argue that this, is a better default value ($\eta_d$) for modelling MCs from the WISE/NEOWISE data catalogue. Since the $\eta$-value distribution reflects the size-related biases inherent to the WISE survey for the MCs, this choice is not necessarily valid for other asteroid groups. 

Although the MCs are a compositionally heterogeneous population \citep[][and references therein]{deLeon2010b}, we do not find any statistical indications of compositional heterogeneity in their $\eta$-value distributions. We took the sample of panel B and separated it into objects with albedos lower than or equal to 0.12 and those higher than 0.12, but we see almost identical histograms (panels C and D in Fig.~\ref{fig:hist_eta}), with KS $p$-values greater than 0.98. As we discuss in the following section (see Fig.~\ref{fig:hist_logpV}), this albedo value roughly distinguishes between C-complex and S-complex spectral classes \citep[although X-types fall in both albedo groups; see e.g. ][]{Tholen1984,Bus2002a,Bus2002b,DeMeo2009,Mainzer2011e}. 
\begin{figure}
  \begin{center}
    \includegraphics[width=0.7\hsize]{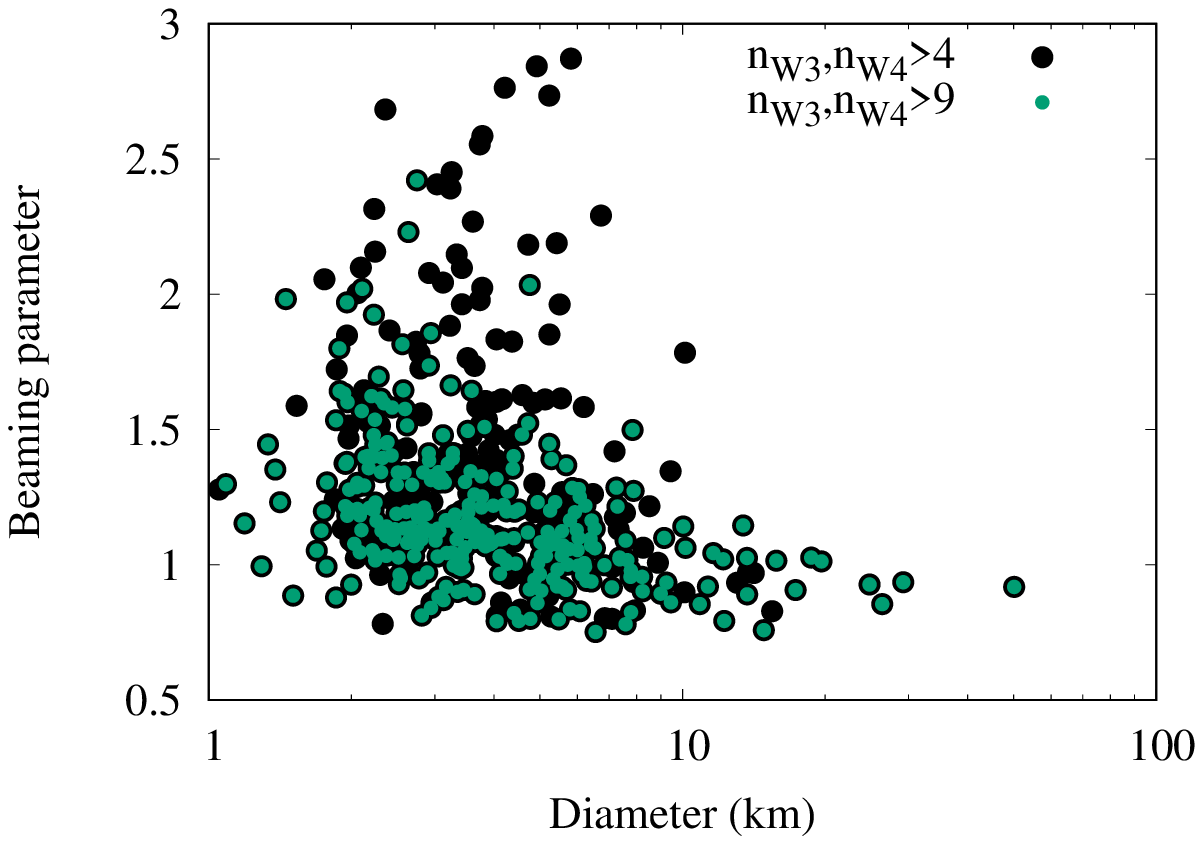}
    \caption{
      Beaming parameter versus size for Mars-crossing asteroids determined from at least five (large black circles) and ten (small green circles) measurements in each purely thermal band. Error bars in diameters ($>$10\%) and beaming parameters ($>$15\%) are not shown for better visibility.
      \label{fig:eta_vs_size}
    }
  \end{center}
\end{figure}


On the other hand, the beaming parameters of the largest bodies do seem to be lower than those of the smaller ones (Fig.~\ref{fig:eta_vs_size}). 
We have 19 objects with $D\geq 10$~km and 267 objects with $D<10$~km among the ``n$_\mathrm{W3}$,n$_\mathrm{W4}>$9'' sample, with average $\eta$ of 1.0~$\pm$~0.1 and 1.2~$\pm$~0.2, respectively. The KS rules out that both samples are drawn from the same distribution with a $p < 10^{-5}$. 
Although this could plausibly be related to the expected presence of finer regolith on the surfaces of larger bodies \citep{Delbo2007,Delbo2009,Delbo2015}, it is not possible to give a purely physical interpretation to this trend without modelling additional physical information because the beaming parameter is not a physical property of the surfaces. Rotational periods\footnote{We have found 322 MCs in our list with reported periods in the Asteroid Lightcurve Database managed by B. Warner (http://www.minorplanet.info/home.html). See \citet{Warner2009}.}, spin pole orientations and shapes are required to use thermophysical models to infer thermal inertias, which reveal information about the physical nature of the surface materials \citep[see][for a review]{Delbo2015}, but work in this direction is far beyond the scope of this paper. 

Moreover, there may be other causes for this trend. For example, it is known that $\eta$ increases with the phase angle of the observations \citep[e.g.][]{Delbo2004,Wolters2009}, and because WISE observed in quadrature, many MCs were detected at relatively short heliocentric distances, i.e., high phase angles reaching up to 50 degrees. This also explains why the average MC beaming parameter is intermediate between that of main belt asteroids and NEAs \citep[see Fig.~7 in][]{Mainzer2011}. 

Also, the lower quality of the data for small objects and the fact that these objects tend to have more irregular shapes \citep[e.g.][]{Harris1979,Nortunen2017} can have an impact on the dispersion of the $\eta$ values, so as we increase the minimum number of data points in bands W3 and W4 we remove most of the more extreme beaming parameters, $\eta>2$ (cf. the black and green circles of Fig.~\ref{fig:eta_vs_size}).

\subsection{Visible geometric albedos \label{sec:albedos}}

\begin{figure}[h!]
  \begin{center}
    \includegraphics[width=0.7\hsize]{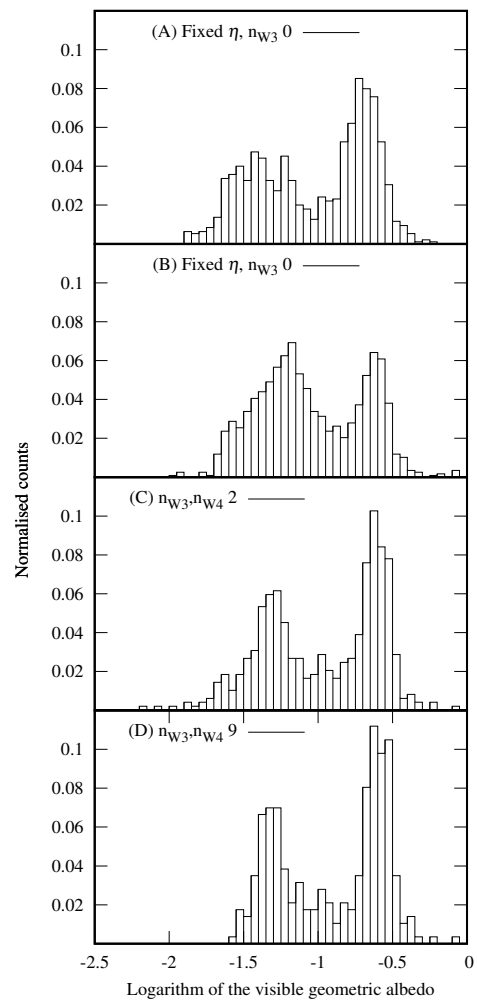}
    \caption{
      Histograms of visible geometric albedos. The upper panels show the samples for which we had to assume a default value of beaming parameter to fit the diameter to W2 data only (A) and W3 data only (B). The lower panels show the samples with diameter and $\eta$ fitted to three or more (C) and ten or more (D) data points in both thermally dominated bands, W3 and W4. 
      \label{fig:hist_logpV}
    }
  \end{center}
\end{figure}

Our sample presents two peaks in the albedo histograms shown in Fig.~\ref{fig:hist_logpV}. These peaks correspond to the C-complex and the low-albedo component of the X-complex on the one hand, and to the S-complex and high-albedo X-complex on the other hand \citep[][and references therein]{Mainzer2011e}. Panel A includes all objects whose diameter was fitted to W2 data only, with fixed beaming parameter ($\eta_d=1.20$; case 3 in Sect.~\ref{sec:model}). Panel B also shows the histogram for those objects with fixed $\eta$, but the diameter is fitted to W3 data (case 2 in Sect.~\ref{sec:model}). Panels C and D include objects with $\eta$ fitted to at least three and ten data points, respectively. 

Comparing the first two panels with the other two illustrates the effect of fixing the beaming parameter to a default value: for the samples with fixed $\eta$, the lower albedo peaks are broader and the histograms are slightly shifted. 
The lower two histograms show how increasing the minimum number of data points available for the fits removes most of the extremely low albedos from the sample with fitted $\eta$. 

It is also apparent how the low-albedo peak in panel B is more populated than the high-albedo peak, whereas the converse is true for the other samples. This could be explained by the fact that band W3 has a higher sensitivity than band W4, which makes it more likely for low-albedo objects to be detected in band W3 alone since their surface temperatures tend to be higher (see Eq.~\ref{eq:energy}). Instead, higher albedo objects are more likely to verify the requirement that both W3 and W4 data should be available.

\subsection{Biases in the sizes and albedos related to the choice of a default value of the beaming parameter\label{sec:uncertainties}}

In this section we discuss how our particular choice of the default beaming parameter ($\eta_d$) affects the corresponding sizes and visible albedos. We examined how modifying the value of $\eta_d$ (by up to $\pm 40$\%) changes the sizes and albedos of a particular object. We chose asteroid (90943) because it was observed in all phases of the WISE survey and thus has several groups of observations verifying the three situations enumerated in Sect.~\ref{sec:model}.

\begin{figure}
  \begin{center}
    \includegraphics[width=0.7\hsize]{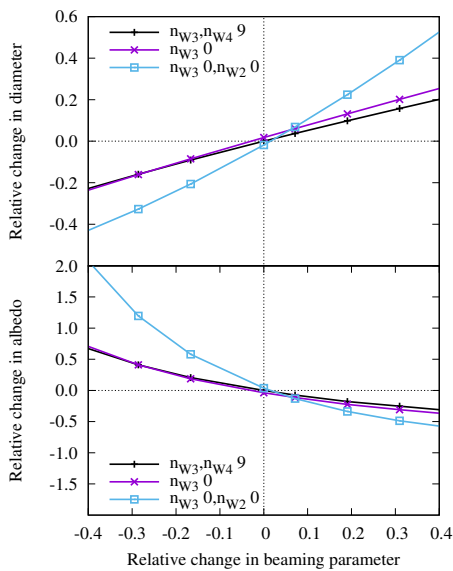}
    \caption{ 
      Relative change in diameter (top) and visible geometric albedo (bottom) as a function of the relative change in the beaming parameter for MC asteroid (90943). We took the fitted value of $\eta$ obtained from the first group of observations (labelled n$_\mathrm{W3}$,n$_\mathrm{W4}>9$)  as reference and fitted diameters using different fixed $\eta$ values (up to $\pm$40\%) for three groups of observations. The label n$_\mathrm{W3}$,n$_\mathrm{W4}>9$ means that W3 and W4 data were available (more than 9 data points in each band), n$_\mathrm{W3}>0$ indicates that the diameter was only fitted to 12 $\mu$m (W3) data, and n$_\mathrm{W3}=0$,n$_\mathrm{W2}>0$ only to 4.6 $\mu$m (W2) data only. We note the different scales in the panels. 
      \label{fig:Delta_eta}
    }
  \end{center}
\end{figure}

The upper panel of Fig.~\ref{fig:Delta_eta} shows the relative change in size as a function of the relative change in $\eta_d$ for asteroid (90943). 
In the first case we have purely thermal data available in both thermal bands (labelled n$_\mathrm{W3}$,n$_\mathrm{W4}>$9), but in the second group of observations we have data in one thermal band only (n$_\mathrm{W3}>$0). If we take the fitted $\eta$-value of the first case (n$_\mathrm{W3}$,n$_\mathrm{W4}>$9) as reference, the relative change in size scales linearly with the relative change in beaming parameter to the point that 20\% change in $\eta$ results in a 10\% change in diameter. As expected, this is consistent with the minimum error bars of 10\% typically quoted for NEATM. 

In the third case, i.e. when there is no purely thermal data available (empty squares in Fig.~\ref{fig:Delta_eta}), there are several likely reasons why the size is more sensitive to inadequate choices of the beaming parameter. First, we had to assume a value for the W2 (4.6 $\mu$m) geometric albedo in order to estimate the reflected light contribution to the measured W2 fluxes, and the assumed value of 1.4 times the visible geometric albedo may not be particularly adequate for this asteroid. Second, we took the value of $\eta$ fitted to the first group of observations (n$_\mathrm{W3}$,n$_\mathrm{W4}>9$) as a reference, but as we have already discussed in Sect.~\ref{sec:model}, this value is not a property of the object and may not be appropriate for modelling other groups of observations\footnote{For example, asteroid (408795) has two groups of observations taken at widely different dates with a similar geometry of observation. However, we obtained almost identical sizes (2.166 versus 2.099 km) but significantly different beaming parameters (1.4 versus 1.13). Conversely, for asteroid (475) Ocllo, observed during the Reactivation phase on three occasions, our default $\eta_d$ results in three different size values, 34 km (39 degrees phase angle), 25 km (27 degrees), and 37 km (18 degrees), which suggests this object can be significantly irregular.}. Thus, different viewing geometries and/or object orientations can account for the greater effect on equivalent size determination seen in this third case. Finally, asteroid fluxes are typically faint in band W2 (4.6 $\mu$m) plus the constant 0.9 emissivity assumption may break down at wavelengths shorter than $~$8~$\mu$m \citep{Delbo2015}. 

\begin{figure}
  \begin{center}
    \includegraphics[width=0.7\hsize]{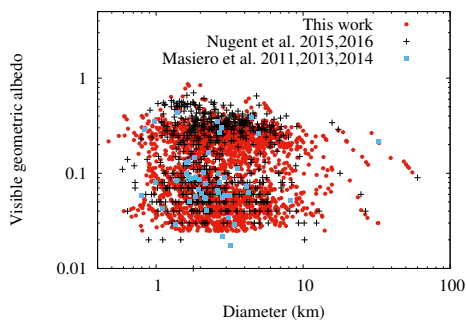}
    \caption{
      Visible geometric albedo versus diameter for a sample of MCs that are featured both in \citet{Masiero2011} or \citet{Nugent2015,Nugent2016} and this work.
      The alignment of some groups of points in horizontal or diagonal lines are artefacts and are discussed in the text. 
      \label{fig:eta_vs_size_comparison}
    }
  \end{center}
\end{figure}

The visible geometric albedo is more strongly and non-linearly affected, which is expected since $p_V \propto D^2$ (lower panel in Fig.~\ref{fig:Delta_eta}). Moreover, underestimating the beaming parameter results in a larger deviation, especially in cases when no W3 or W4 data are available, which also argues in favour of using a higher default value of $\eta_d=1.2$. 
We thus expect our choice of $\eta_d=1.2$ to increase our sizes by 15 to 25\% and decrease our albedos by 30 to 50\% on average with respect to those of the NEOWISE team. To show the effect, we plotted the albedos of the NEOWISE team MCs and ours as a function of diameter in Fig.~\ref{fig:eta_vs_size_comparison}. Indeed, although the two clouds of points (related to the different taxonomies, cf.~Fig.~\ref{fig:hist_logpV}) seem to overlap reasonably  well, our diameters do tend to be somewhat larger and our albedos lower to the extent we expected. 

There are two artefacts apparent in Fig.~\ref{fig:eta_vs_size_comparison} that deserve being mentioned. On the one hand, the Nugent et al. points in the low-albedo cloud are aligned in horizontal lines because their published $p_V$ values are rounded off. In our case, we can see streaks of points aligned diagonally in the less populated regions. These correspond to curves of constant $H$ (straight lines in our log-log plot). In particular, some of these aligned points belong to the same objects, those with more than one entry in our catalogue. Most noticeably, the group of points to the right of the plot correspond to asteroid (132) Aethra, the largest object in our sample, which was observed several times in post-cryogenic phases. 

The above demonstrates how the errors in the diameters (albedos) derived from NEATM with a fixed $\eta$-value can be significantly larger than the often quoted minimum values of 20\% (40\%) for these cases \citep{Mainzer2011b,Masiero2012,Nugent2015} if the chosen default value $\eta_d$ is not appropriate, especially for very small objects observed at high phase angles. 
Although \citet{Nugent2016} compared the diameters of 23 objects in their catalogue for which ground-truth values are available, this sample is too small to provide a definitive estimate of size and albedo error bars since they do not represent all asteroid populations. However, a better examination of how these factors influence the NEATM diameters would require many more ground-thruth values, asteroid shapes that are currently unavailable, and better statistics of the distribution of asteroidal spin vector orientations \citep{Hanus2011}.  
Thus, for the moment, we emphasise that the quoted uncertainties are only minimum values and that caution is required when analysing diameters and albedos derived from NEATM.

\section{Conclusions}

We provide a set of equivalent diameters and visible geometric albedos ($p_V$) of Mars-crossing asteroids derived from WISE/NEOWISE data. We fitted the infrared beaming parameter ($\eta$) of more than 400 MCs observed during the fully cryogenic part of the mission, most of which have not been published previously. We also report diameters and albedos of 1572 MCs, 949 observed only in one thermal band (W3, 12 $\mu$m), and 891 observed in band W2 (4.6 $\mu$m). Our results are collected in Table~\ref{table:all} (full version available electronically). 

We compared our diameters and albedos with those of MCs featured in \citet{Masiero2011} or \citet{Nugent2015,Nugent2016} and showed that we obtain similar results when the input parameters are the same, including their default value of the beaming parameter when it could not be fitted. More specifically, our sizes tend to be about 2\% lower and our albedos 3\% higher. These offsets are well within the typical errors expected for the NEATM (minimum of 10\% for diameter and 20\% for albedos). 

The $p_V$-value distribution shows two peaks, one at $\sim$0.06, the other at $\sim$0.26. This is expected since the MCs are compositionally diverse and include members of all the main spectroscopic classification groups. 

The $\eta$-value distribution, on the other hand, does not reflect this compositional heterogeneity. It peaks at a value of 1.20 (see Figs.~\ref{fig:hist_eta} and \ref{fig:eta_vs_size}), which is higher than the average $\eta$ of main belt asteroids \citet{Masiero2011} but lower than for NEAs \citep{Mainzer2011}. 
This is actually expected for a population with intermediate semi-major axes, since the beaming parameter is known to correlate with phase angle and WISE observed in quadrature, which means that objects closer to the sun were observed at higher phase angles. 

In that sense, because the beaming parameter is not a physical property of the asteroids and given the inevitable effects of size-related biases, we argue that  the average $\eta$ of the population is the best default value of the beaming parameter ($\eta_d$), which also means this choice is not generally valid for other asteroid groups observed by WISE. We thus remark that the features mentioned here only reflect the properties of the MCs within the limits of the size-related biases inherent to the WISE survey and that more in-depth studies would require an assessment of the completeness of the catalogue \citep[][and references therein]{Mainzer2011,Mainzer2014b}.

Finally, since we are proposing a different value of $\eta_d$ than the NEOWISE team for the MCs, in Sect.~\ref{sec:uncertainties} we examined how the diameters and albedos are affected. Our higher $\eta_d$ produces  15 to 20\% larger diameters and therefore 30 to 50\% lower albedos on average (see Fig.~\ref{fig:eta_vs_size_comparison}).

\begin{acknowledgements}

  The results described here have been made possible by the NEOShield-2 project, which has received funding from the European Union's Horizon 2020 research and innovation programme under grant agreement no. 640351.
  
  The research leading to these results has received funding from the European Union’s Horizon 2020 Research and Innovation Programme, under Grant Agreement no. 687378.
  
  This research has made use of data and/or services provided by the International Astronomical Union's Minor Planet Center.
  
  This publication makes use of data products from NEOWISE, which is a project of the Jet Propulsion Laboratory/California Institute of Technology, funded by the Planetary Science Division of the National Aeronautics and Space Administration.
  
  This research also made use of the NASA/ IPAC Infrared Science Archive, which is operated by the Jet Propulsion Laboratory, California Institute of Technology, under contract with the National Aeronautics and Space Administration.  

This research has made use of IMCCE's Miriade VO tool.
  
\end{acknowledgements}

\bibliographystyle{aa} 
\bibliography{AsteroidsGeneral}

%
%

\end{document}